# Medial Residual Encoder Layers for Classification Brain Tumors in Magnetic Resonance Images

Zahra Sobhaninia[1], Nader Karimi[1], Pejman Khadivi[2], Shadrokh Samavi[1,3]
[1]Isfahan University of Technology, Isfahan, 84156-83111 Iran,
[2]Computer Science Department, Seattle University, Seattle, 98122 USA
[3]Elect. & Comp. Engineering, McMaster University, L8S 4L8, Canada

*Abstract*— **According to the World Health Organization (WHO), cancer is the second leading cause of death worldwide, responsible for over 9.5 million deaths in 2018 alone. Brain tumors count for one out of every four cancer deaths. Therefore, accurate and timely diagnosis of brain tumors will lead to more effective treatments. Physicians classify brain tumors only with biopsy operation by brain surgery, and after diagnosing the type of tumor, a treatment plan is considered for the patient. Automatic systems based on machine learning algorithms can allow physicians to diagnose brain tumors with noninvasive measures. To date, several image classification approaches have been proposed to aid diagnosis and treatment. For brain tumor classification in this work, we offer a system based on deep learning, containing encoder blocks. These blocks are fed with post-max-pooling features as residual learning. Our approach shows promising results by improving the tumor classification accuracy in Magnetic resonance imaging (MRI) images using a limited medical image dataset. Experimental evaluations of this model on a dataset consisting of 3064 MR images show 95.98% accuracy, which is better than previous studies on this database.**

## I. INTRODUCTION

Brain tumor type diagnosis is the first and vital step to consider an effective plan to treat this disease. So accurate and prompt diagnosis has a significant role in this situation. These days, computer-aided diagnosis (CAD) techniques have been helping physicians in extensive applications of brain tumor analysis such as tumor detection, classification, localization, and segmentation. In this regard importance of detection of brain tumors has led to the proposal of various automated and semi-automatic methods. [1] [2] [3] [4].

Various methods are proposed for image classification, extracting features from the raw input data, and applying classification algorithms on the extracted features. For instance, for brain tumor classification, Cheng et al. [3] extracted the region of interest (ROI) by using morphological operation. Then, they used a space pyramid matching (SPM) method for better feature extraction. Extracted features were given to some classifiers such as SVM, SRC, and KNN to perform the classification operation [5]. In another approach, Asodekar et al. [6], after shape-based features extraction, applied SVM and random forest classifiers for the classification task.

A convolutional neural network (CNN) is a machine-learning algorithm that significantly affects image segmentation and classification without handcraft feature extraction. For example, Abiwinanda et al. [7] applied CNN for tumor classification. Ismael et al. used a CNN network on MRI images and statistical image features as input data for tumor classification [8]. Afshar et al. [9] studied a different architecture of CapsNet networks for tumor type classification. Also, Pashaei et al. [2] and Gumaei et al. [1] presented a method based on deep ELM networks. In another work, Ghassemi et al. applied generative adversarial networks (GANs) for brain tumor classification [10].

This paper proposes an approach based on a deep neural network called the ResBlock classifier to classify brain tumor types in MRI images. The database studied in this research is more extensive than other available databases on brain tumors. However, it is not an ideal database for training deep learning approaches in terms of the number of images. Besides providing a high-precision deep learning network approach, we intend to present the network's architecture as simple as possible to cope with a limited medical database. We have focused on a brain tumor database consisting of 3064 T1-contrast MRI images in this research. We don't apply data augmentation, unlike previous works on this database. The experimental results show that the proposed model is competitive with state-of-the-art methods with data augmentation, although we have not used data augmentation.

The structure of the paper is as follows. In section 2, we will discuss the background of the research. Then, the proposed method is presented in section 3, and experimental results are discussed in section 4 of the paper. Finally, section 5 is dedicated to concluding remarks.

## II. BACKGROUND AND PROBLEM FORMULATION

The brain is one of the most important and complex human organs protected inside the skull. This mass of nervous tissue can suffer tumors due to abnormal and uncontrolled cell growth. In general, tumors can be divided into three categories: benign, pre-carcinoma, or malignant. The difference between benign and malignant is that benign tumors do not invade other tissues and organs of the body and can surgically be removed [11]. More specifically, primary tumors are gliomas, meningioma, and pituitary tumors. One of the differences between these tumors is that generally, gliomas are malignant, while meningioma tumors are typically benign. Slow-growing tumors are benign and malignant, but even benign tumors can damage other organs [12]. An essential step in effectively treating tumors is early diagnosis and accurate identification of their types. MRI is a non-ionizing medical imaging modality that provides better contrast in the body's soft tissues.

Furthermore, MRI distinguishes better between muscle, fat, water, and other soft tissues. MRI is the most prevalent human brain tumor technique that allows experts to diagnose tumor presence and type. However, correct diagnosis depends on the experience and expertise of specialists investigating the characteristics of the images. Also, these analyses are time-consuming and prone to error. In this regard (CAD) can be used to help tumor diagnostics without surgery or invasive methods. Up to now, all CAD approaches can be divided into two groups. The first group of CAD methods can classify images with brain tumors and those with no tumors. Algorithms of the second category of CAD methods classify brain images depending on brain tumor pathological types. The distinction between normal and abnormal images is relatively easy and can be performed by extracting handcrafted features. In contrast, identifying the tumor type is challenging because of varius shapes, sizes, and textures of brain tumors. Some samples of MR images that contain brain tumors are shown in Fig 1.

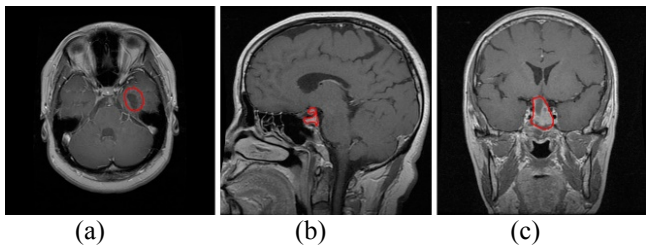

(a)  (b)  (c)

Fig 1; Samples of brain MR images contain three common tumors (a) glioma, (b) meningioma, and (c) pituitary tumor. Tumors specified with red boundary lines [13]

Accurate classification of tumor types is not possible by handcrafted features, and deep neural networks should be used for this purpose. Another difficulty in the classification of tumor types is the presence of different MR images captured from different views. For example, in Fig. 1, we can see (a) axial, (b) sagittal, and (c) coronal images are taken from three directions. Our proposed network overcomes the mentioned challenges.

## III. PROPOSED METHOD

Generally, primitive CNNs used for classification tasks have similar structures and are only different in the number of convolution layers, stride, and kernel dimensions [14]. The proposed CNN is inspired by LeNet [15] and is used to classify images according to the tumor type. We first modified LeNet to improve the accuracy by some changes in the arrangement and number of layers. As shown in Fig 2, this network consists of 18 convolution and pooling layers that extract features from raw images. After features are extracted with passing from these layers, two fully connected layers are applied with 4096 nodes. Extracted features are sent to these fully connected layers, and tumor type classification is performed in the last layer. The final fully connected layer has three neurons representing glioma, meningioma, and pituitary tumors.

After evaluating this CNN network performance and observing satisfactory results of brain tumor classification on dataset [13], another CNN network is proposed that is inspired by LeNet architecture and residual learning [16].

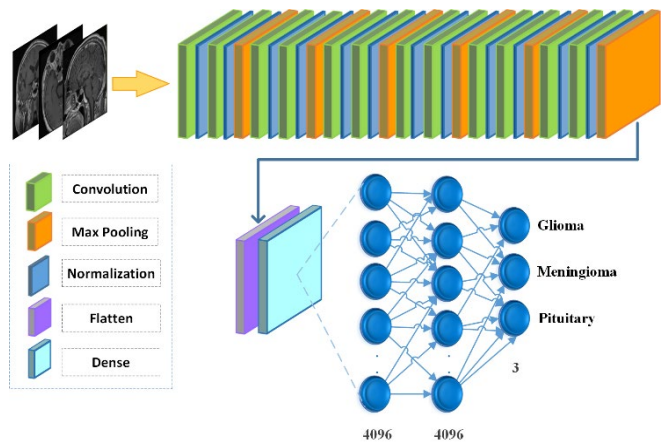

Fig 2: The proposed LeNet inspired *[15]* model

Instead of considering sequential convolutional and pooling layers in the proposed classification network, we propose an architecture containing some encoders. The encoder blocks called medial residual blocks (MidResBlock) are considered for encoding data. These blocks consist of some sequential convolutional layers, a normalization layer, a max-pooling layer, and a skip connection between some of them. The structure of the proposed encoding block, MidResBlock, is shown in Fig. 3.

Applying sequential convolution and pooling layers in large numbers in the elementary CNN networks and passage of information through these layers lead to the loss of some desirable features. This loss of features reduces the performance and accuracy of the network. For this reason, the residual link is considered in the structure of each encoder block. As observed in Fig. 3, after passing data from the first few layers, the output feature map is smaller than the input of the block because of the max-pooling layer. Hence, we use residual learning by adding medial features and the encoder's output features with the same size.

Furthermore, the position of the residual link is important to reduce losing extraction information; in this regard, we have implemented different positions of residual link in the encoder block. For instance, considering two residual links in the encoder block, a residual link between input and the medial data and also a residual between medial data and output of encoder block. Eventually, the best result is obtained from the presented model. These blocks provide a network architecture model to monitor its behavior for brain tumor classification better.

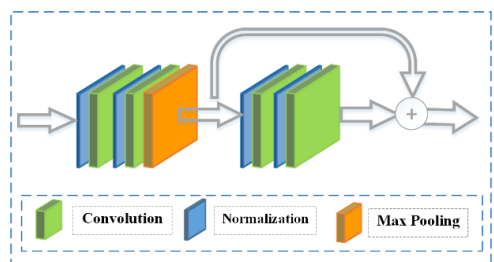

Fig 3: Proposed MidResBlock encoder block

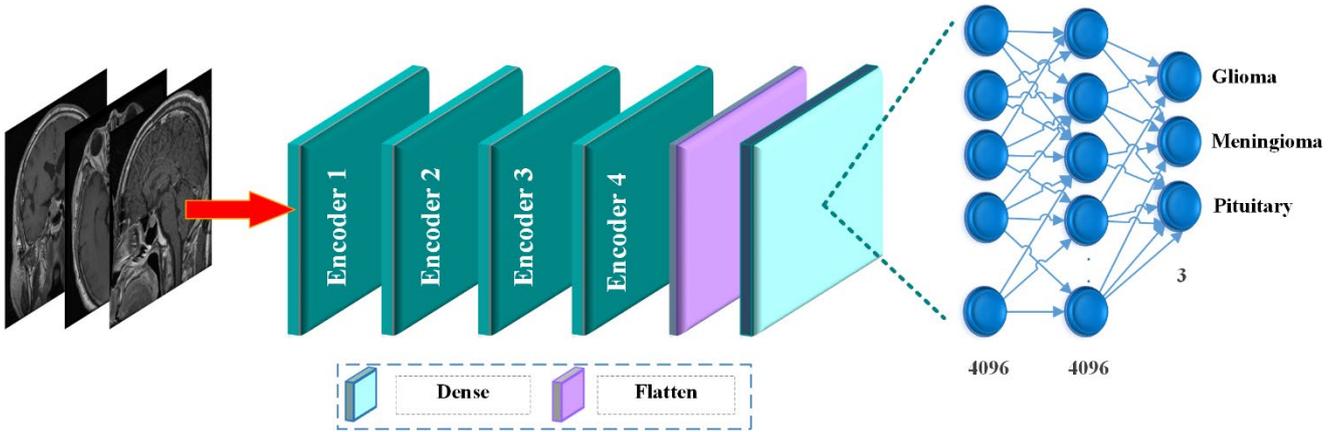

Fig 4: Proposed classifier using MidResBlocks (MidResBlock classifier network)

The experimental results show that Adding information of middle convolution layers to each encoder layer's output helps preserve information during the training process compared to other structures mentioned above. The proposed network structure is shown in Fig 4. This network, called the MidResBlock classifier network, consists of four MidResBlock as an encoder unit that extracts features. The depth of deep networks affects the performance model; in this regard, in this work, the number of encoders for feature extraction is optimal based on the complexity of the data set and the limited number of images. Extracted features are passed to two fully connected layers, determining dimensions with 4096 neurons. Then the network performs the classification operation with the last layer, which contains three neurons representing tumor types. We applied categorical cross-entropy for the proposed network training as the loss function that is most commonly used for multi-label classifications.

## IV. EXPERIMENTAL RESULTS

Presented models were trained and evaluated on a benchmark brain tumor MRI images dataset [13], containing 3064 T1-weighted MR images gathered from 2005 to 2010. Three experienced radiologists have manually bordered tumor areas. These images include 1426 gliomas, 930 pituitary, and 708 meningioma tumors. The resolution of images is 512×512, with 0.49×0.49 mm2 pixel size. Our model is implemented on Python 3.7 and Tensorflow. Proposed models trained over 150 epochs using stochastic gradient descent (SGD algorithm) with momentum (learning rate = 0.001). Training time on NVIDIA GeForce GTX 1080 Ti was about 12 hours.

The Data is divided to demonstrate the uniform distribution of tumor types based on the cross-validation method to address the imbalanced dataset to evaluate our model. In this paper, mean accuracy is used to report the final result. We used k-fold learning; hence the average of all folds is reported. Accuracy, a typical criterion used to evaluate classification models, is defined as the ratio of correct predictions by the total number of predictions. This criterion works poorly for an imbalanced dataset, as it may report high accuracy while biased toward the class with the largest number of samples.

In addition, the confusion matrix is used to show precision criteria using the misclassified samples. The confusion matrix determines that which samples are classified in the wrong classes. *Table 1* shows the confusion matrix of the predicted tumor type of our model.

Table 1: Confusion matrix of predicted tumor type by considering Aggregation Module. The tumor type precision is 96.08%, which is the average Precision of each tumor type

| True/Predicted | Glioma | Pituitary | Meningioma | **Precision** |
|---|---|---|---|---|
| Glioma | 1375 | 24 | 21 | 96.987% |
| Pituitary tumor | 22 | 887 | 17 | 95.787% |
| Meningioma | 23 | 15 | 680 | 94.507% |

The accuracy of the proposed model for brain tumor classification is investigated. Table 1 shows the results of the two proposed networks of LeNet-based and the MidResBlock classifier. Also, the results of references [4] and [5], using the same benchmark data set, are shown in Table 1. Pashaei et al. [2] report that they achieved an accuracy of 93.68% by presenting a deep ELM network method. Gumaei et al. [1] reported 94.23% accuracy using a feature extraction method and a deep ELM network. The reported evaluation based on the accuracy indicates that considering the residual links in the network classification architecture positively affects the classification accuracy. The confusion matrix of the proposed model is illustrated inTable 2.

Table 2: Quantitative results of brain tumor classification

| Proposed Method | Accuracy % |
|---|---|
| Proposed LeNet-based network | 90.06 |
| Pashaei et al. [2] | 93.68 |
| Gumaei et al. [1] | 94.233 |
| generative adversarial networks [10] | 95.6 |
| Proposed MidResBlock classifier network | **95.98** |

## V. CONCLUSION

Providing accurate diagnostic tools for detecting brain tumors can significantly impact early cancer detection, treatment, and outcome. This paper presented a classification architecture with high accuracy for detecting brain tumors from MRI images. Our results show that adapting network architecture by modifying the number of layers and adding residual links as skip connections has impressive classification accuracy. Given the limitation of medical image datasets, CNN's appropriate depth and simplicity significantly improve accuracy. Our results show that the proposed classifier using MidResBlock encoder layers is more accurate than the state-of-the-art. Furthermore, due to model simplicity, it can preserve important image features and information. Our future work involves applying the proposed model to other medical images. We intend to use the proposed method for ultrasound [17], histopathological [18], non-dermoscopic images [19], OCT images [20], abdominal CT images, and vessel segmentation in angiograms [22].